\documentclass{aastex63}

\def \UMPhysics {Department of Physics, University of Michigan, Ann Arbor, MI 48109, USA}
\def \UMAstronomy {Department of Astronomy, University of Michigan, Ann Arbor, MI 48109, USA}
\def \UW {DiRAC Institute and the Department of Astronomy, University of Washington, Seattle, USA}

\usepackage{CJKutf8}
\received{}
\revised{}
\accepted{}
\submitjournal{AJ}

\shorttitle{DEEP IV: Single-exposure TNOs}
\shortauthors{Strauss et al.}
\graphicspath{{./}{}}
\begin{document}

\title{The DECam Ecliptic Exploration Project (DEEP) IV:\\ Constraints on the shape distribution of bright TNOs}

\def \NAU {Department of Astronomy and Planetary Science, Northern Arizona University,\\PO Box 6010, Flagstaff, AZ 86011, USA}

\def \UMPhysics {Department of Physics, University of Michigan, Ann Arbor, MI 48109, USA}

\def \UMAstronomy {Department of Astronomy, University of Michigan, Ann Arbor, MI 48109, USA}

\def \UW {DiRAC Institute and the Department of Astronomy, University of Washington, Seattle, USA}

\def \uchile {Departamento de Astronomía, Universidad de Chile,\\ Camino del Observatorio 1515, Las Condes, Santiago, Chile}

\def \cfa {Harvard-Smithsonian Center for Astrophysics,\\ 60 Garden St., MS 51, Cambridge, MA 02138, USA}

\def \byu {Department of Physics and Astronomy, Brigham Young University, Provo, UT 84602, USA}

\def \apl {Applied Physics Lab, Johns Hopkins University,\\ 11100 Johns Hopkins Road, Laurel, Maryland 20723, USA}

\def \ucla {Department of Earth, Planetary and Space Sciences, University of California Los Angeles, 595 Charles E. Young Dr. East,
Los Angeles, CA 90095, USA}

\def \carnegie {Earth and Planets Laboratory, Carnegie Institution for Science, Washington, DC 20015}

\def \stgallen {School of Computer Science, University of St. Gallen,\\ Rosenbergstrasse 30, CH-9000 St. Gallen, Switzerland}

\correspondingauthor{Ryder Strauss}
\email{rstrauss@nau.edu}

%%%%%%%%%%%%%%%%%%%%%%%%%%%%%%%%%%%%%%%%%%%%%%%%%%%%%%%%%%%
\author[0000-0001-6350-807X]{Ryder Strauss}
\affiliation{\NAU}

\author[0000-0003-4580-3790]{David E. Trilling}
\affiliation{\NAU}

\author[0000-0003-0743-9422]{Pedro H. Bernardinelli}
\affiliation{\UW}

\author{Christiano Beach}
\affiliation{\UMPhysics}

\author[0000-0001-5750-4953]{William J. Oldroyd}
\affiliation{\NAU}

\author[0000-0003-3145-8682]{Scott S. Sheppard}
\affiliation{\carnegie}

\author[0000-0002-0298-8089]{Hilke E. Schlichting}
\affiliation{\ucla}

\author[0000-0001-6942-2736]{David~W.~Gerdes}
\affiliation{\UMPhysics}
\affiliation{\UMAstronomy}

\author[0000-0002-8167-1767]{Fred C.~Adams}
\affiliation{\UMPhysics}
\affiliation{\UMAstronomy}

\author[0000-0001-7335-1715]{Colin Orion Chandler}
\affiliation{LSST Interdisciplinary Network for Collaboration and Computing, 933 N. Cherry Avenue, Tucson AZ 85721}
\affiliation{\NAU}

\author[0000-0002-5211-0020]{Cesar Fuentes}
\affiliation{\uchile}

\author[0000-0001-8550-6788]{Matthew J. Holman}
\affiliation{\cfa}

\author[0000-0003-1996-9252]{Mario Juri\'c}
\affiliation{\UW}

\author[0000-0001-7737-6784]{Hsing~Wen~Lin (\begin{CJK*}{UTF8}{gbsn}
林省文\end{CJK*})}
\affiliation{\UMPhysics}

\author[0000-0002-2486-1118]{Larissa Markwardt}
\affiliation{\UMPhysics}

\author{Andrew McNeill}
\affiliation{\NAU}
\affiliation{Department of Physics, Lehigh University, 16 Memorial Drive East, Bethlehem, PA, 18015, USA}

\author[0000-0002-7817-3388]{Michael Mommert}
\affiliation{\stgallen}

\author[0000-0003-4827-5049]{Kevin J. Napier}
\affiliation{\UMPhysics}

\author[0000-0001-5133-6303]{Matthew J. Payne}
\affiliation{\cfa}

\author[0000-0003-1080-9770]{Darin Ragozzine}
\affiliation{\byu}

\author[0000-0002-9939-9976]{Andrew S. Rivkin}
\affiliation{\apl}

\author[0000-0002-7895-4344]{Hayden Smotherman}
\affiliation{\UW}

\author[0000-0001-9859-0894]{Chadwick A. Trujillo}
\affiliation{\NAU}



\begin{abstract}

We present the methods and results from the discovery and photometric measurement of 26 bright (VR $>$ 24 trans-Neptunian objects (TNOs) during the first year (2019-20) of the DECam Ecliptic Exploration Project (DEEP). The DEEP survey is an observational TNO survey with wide sky coverage, high sensitivity, and a fast photometric cadence.  We apply a computer vision technique known as a progressive probabilistic Hough transform to identify linearly-moving transient sources within DEEP photometric catalogs. After subsequent visual vetting, we provide a photometric and astrometric catalog of our TNOs. By modeling the partial lightcurve amplitude distribution of the DEEP TNOs using Monte Carlo techniques, we find our data to be most consistent with 
an average TNO axis ratio  b/a $<$ 0.5, implying a population dominated by non-spherical objects. Based on ellipsoidal gravitational stability arguments, we find our data to be consistent with a TNO population containing a high fraction of contact binaries or other extremely non-spherical objects. We also discuss our data as evidence that the expected binarity fraction of TNOs may be size-dependent. 

\end{abstract}

\keywords{Trans-Neptunian objects --- Surveys ---
Small Solar System bodies ---
Solar System ---
Kuiper Belt --- Photometry}

\section{Introduction}\label{sec:intro}

The outer reaches of the Solar System are home to an extremely primitive population of small objects. These trans-Neptunian objects (TNOs), named such because they orbit the Sun at the distance of Neptune and beyond, preserve a fossil record from the earliest stages of Solar System formation. In studying these objects, we can obtain enormous insight into the conditions within the protoplanetary disk, and the mechanisms by which the Solar System evolved into its current state. 

Effective constraints on the shape distribution of TNOs are extremely useful for constraining formation mechanisms. For example, several formation models introduce a preference for forming binaries, and in turn, contact binaries, within the outer solar system \citep{NESVORNY2019, MCKINNON2020,Nesvorny2010,Nesvorny2018,Fraser2017,SHOWALTER2021}. Characterizing the fraction of possible contact binaries within this region provides good observational constraints on the validity of such formation models. 

Confirming an outer Solar System object as a contact binary is extremely difficult though - only the largest TNOs are possible to spatially resolve, even with our best diffraction-limited telescopes \citep{Brown2004,Brown2006}. More sophisticated measurement techniques provide powerful tools for addressing this difficulty and studying TNOs in greater detail. For example, Cold classical TNO 486958 Arrokoth was identified as a likely contact binary long before its shape was confirmed by the spacecraft encounter, thanks to an occultation of a distant star \citep{STERN2019,Buie2020}.Rotational lightcurves can also provide insight into an unresolved object's shape - as an irregular body rotates, its cross-section in the sky changes, as does the amount of reflected solar flux as measured by a distant observer. The shape of the measured lightcurve can be used to constrain the shape of the object itself. 

As of 2022, fewer than 5,000 TNOs were listed in the Minor Planet Center database (https://minorplanetcenter.net/data). Of these, very few TNOs have published lightcurves \citep{SHOWALTER2021,Audrey_2019,WARNER2009}, and even fewer have lightcurves resolved enough to be identified as contact binaries. This determination is commonly made based on details about the lightcurve's shape: a rotating contact binary will often produce a rotational lightcurve with soft, u-shaped peaks and sharp, v-shaped troughs - distinct from the symmetrical sinusoidal lightcurve of a single rotating ellipsoid \citep{Sheppard2002,Lacerda2011,Thirouin2017,ThirouinShep2017,Thirouin_2018,SHOWALTER2021}. 

Because observational measurements of TNOs are so challenging to obtain, most TNO lightcurves have insufficient temporal resolution or signal-to-noise ratio (SNR) to actually identify morphologies indicative of certain object shapes. Instead, the overall lightcurve amplitude distribution can be used as a proxy for object shape distribution. This is the approach taken by \citet{SHOWALTER2021} in their statistical review of TNO lightcurves within the literature. They provide the most complete compilation of published TNO lightcurves to-date, and provide a discussion of their lightcurve amplitude distribution in the context of the shapes distribution of objects within the Kuiper Belt.

The DECam Ecliptic Exploration project (DEEP) is a new observational survey of the Solar System, and is optimized to detect, discover, and characterize TNOs. This survey makes use of the Dark Energy Camera (DECam), a wide-field CCD imager mounted on the 4-meter Victor M. Blanco telescope at Cerro Tololo Interamerican Observatory (CTIO), Chile. DECam instrument details are provided in \citet{DECAM}. Across the first year of the survey alone, DEEP imaged more than 40 unique square degrees of sky, detecting objects as faint as 24th mangnitude in the VR band within individual exposures, and VR $\sim$ 28.23 using digital tracking techniques \citep{Napier_2023}. The details of the DEEP survey design and science goals are discussed in detail in \citet{Trilling_2023}.

DEEP's survey cadence, described in depth in \citet{Trujillo_2023}, makes it very effective at characterizing rotational photometric variability. For each DECam pointing, we image in the VR filter for a series of, nominally, 100 two-minute exposures, giving a total cadence of one image per 2.33 minutes, across about four hours. This fast cadence permits us sensitivity to variations on extremely short timescales with excellent temporal resolution. 

The average TNO lightcurve period is thought to be $\sim$8 hours (Sheppard et al. (2008), Duffard et al. (2009), Thirouin et al. (2010, 2014), Benecchi \& Sheppard (2013)), so the majority of our TNO lightcurves will be partial lightcurves. However, a rotational lightcurve due to an elongated object or contact binary will show two peaks and two troughs, a 4-hour sample of an 8-hour lightcurve will capture the full amplitude of the lightcurve, and the full amplitude can even be captured in as little as 2 hours. For surface albedo variations this is not the case, as there may only be one peak per rotation. Surface heterogeneity not likely to be a strong contributor to the measured light curve amplitude. For example, the Pluto-Charon system is known to have considerable heterogeneity on the surface of both bodies, but the system only has a total photometric variation of only 0.12 amplitudes \citep{BENECCHI2018}. The overall effect of this is that our lightcurve amplitude distribution is likely dominated by shape-driven effects rather than surface heterogeneity. It is thought, however, that many TNOs do have periods well in excess of this average \citep{Kecskemethy2023}. Were we to assume an average rotation period of 20-30 hours instead, the magnitude of the underestimate would increase significantly, meaning the true amplitudes would be far larger than shown here. For this work, we model a wide range of rotation periods to account for this.  Despite measuring primarily partial lightcurves, we do expect that for the majority of our DEEP TNOs, we have captured the full peak-to-trough rotational lightcurve amplitude, and may use that as a reasonable proxy for the object's elongation. 

In this work, we present data processing techniques used to identify bright TNOs within our survey data, as well as the science results of those detections. Section
\ref{sec:dataproc} describes the methods for processing the survey data and identifying the TNOs in that data. Section \ref{sec:results} presents the results of that data processing, Section \ref{sec:discuss} provides a discussion of the results, and Section \ref{sec:conclusions} summarizes conclusions drawn from this work. The computationally expensive digital tracking procedure used to detect the fainest TNOs within the DEEP images is described in \citet{Napier_2023}. Here we discuss results for our bright-end TNOs: those appearing at SNR$\geq$5 in the individual two minute DEEP exposures. 

%%%%%%%%%%%%%%%%%%%%%%%%%%%%%%%%%%%%%%%%%%%%%%%
\section{Data Processing} \label{sec:dataproc}

Once the data have been collected and processed through the DECam Community Pipeline \citep{Valdes2014}, significant additional data processing is required to produce a catalog of lightcurves for analysis. The individual images are downloaded, and for each image the Source Extractor \citep{SEXtractor} tool is used to generate a rudimentary catalog of photometric sources in three dimensions: Right-Ascension, Declination, and time. The same is done for the nightly coadd, which is a single composite image made by projecting all individual images onto the same coordinate system, and combining the images to create one image with much better signal \citep{Zackay2017}. The effect of this is to increase the signal of the stationary sources (those that do not change positions between exposures) and eliminate the transient sources (those that do). The sources in the individual image catalogs are then compared to the stationary sources from the coadded image, and all sources which appear within 0.5 seconds-of-arc of a stationary source are rejected. This does lead to some transient source rejection, but the presence of a star this close to the moving object would likely contaminate those transient sources anyway. The remaining catalogs are dominated by transient sources - such as those of solar system objects moving through the field. 

\begin{figure}
\includegraphics[width=\textwidth]{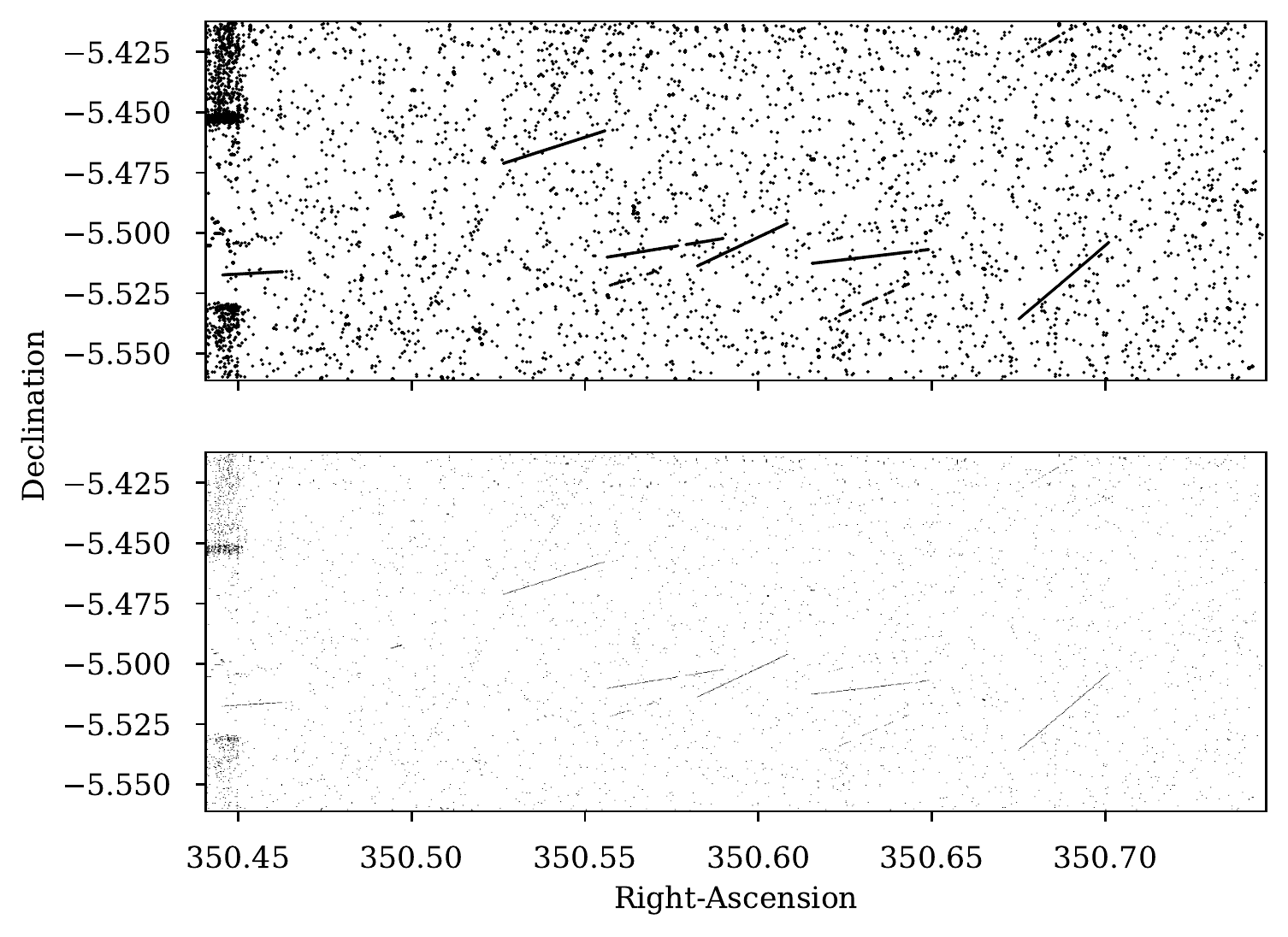}
\caption{Top: Transient catalog for one DECam chip over one four-hour stare. All transient sources for all $\sim 100$ images are shown as black dots. Visually, the linear streaks throughout the field are quite apparent. The overdensity on the left region is an effect of field drift over the four-hour exposure time. Bottom: The same catalog converted into an image format as preparation for the Hough transform streak detection.}
\label{fig:trans} 
\end{figure}

These catalogs consist of tables with each transient source, including the celestial position, apparent magnitude, exposure ID, Modified Julian Date of exposure, and a number of other parameters from Source Extractor. Not every transient source, however, is necessarily from a moving solar system body. Many cosmic rays appear over the course of our four-hour stares, as do artifacts such as those from stellar diffraction spikes, chip edges, or satellites. The next steps in the data processing pipeline are to differentiate moving object sources from those of other artifacts. Over the four-hour time baseline of a DEEP observation, the motion of main-belt asteroids and TNOs are well approximated as linear in right-ascension and declination; so any moving asteroid or TNO within a DEEP stare will manifest as a linearly moving streak within the otherwise fairly random scatter of the transient source catalogs (Figure \ref{fig:trans}). 

The data volume is too large to identify each streak by hand, so instead we used an algorithmic approach to extract the streaks from the object catalog. For this step in the processing pipeline, we made use of the progressive probabilistic Hough transform probhough. This computer vision technique is most commonly used to identify straight edges within images, but it is relatively straightforward to adapt it for streak detection within our transient catalogs. 

The remainder of the analysis is done on a chip-by-chip basis, which limits memory use and allows us to efficidently parallelize our code, as data from each of the 61 chips on DECam can be processed independently of each other. This does not introduce any problems for TNO detection because the rate-of-motion of TNOs is not sufficient to necessitate multi-chip detection. An individual DECam CCD has a field-of-view of roughly 19x8 arcminutes, so even the fastest TNOs will take eight days to move across a chip. TNOs move so slowly relative to the chip gap size (between 40 and 53 arcseconds) that, if one were to trail off a chip edge, we do not expect it to appear on another chip within the four-hour observation. 

Because it is used for computer vision, the Hough transform works most natively on images. To prepare our catalogs for streak detection, we must transform the (ra,dec) catalogs into a 2-dimensional image array. This is done by creating a two-dimensional density histogram in (RA, Dec). The bin sizes in the histogram are very small - about 0.1 arcseconds, considerably smaller than typical atmospheric seeing at CTIO. For comparison, the typical seeing in our dataset ranges roughly from 0.5 arcseconds to 2 arcseconds, with an median of 1.2 arcseconds.  The result is a two-dimensional array of pixels, in which the densest regions of the catalog have the highest pixel values (Figure \ref{fig:trans}). The probabilistic Hough transform is performed on this image array, and a number of (x,y), (x,y) coordinate pairs are returned denoting the pixel positions of the endpoints of all detected lines. These line segments are transformed back into (RA, Dec.) space so that the remainder of the processing can be done with the original source catalogs.

\begin{figure}
\includegraphics[width=\textwidth]{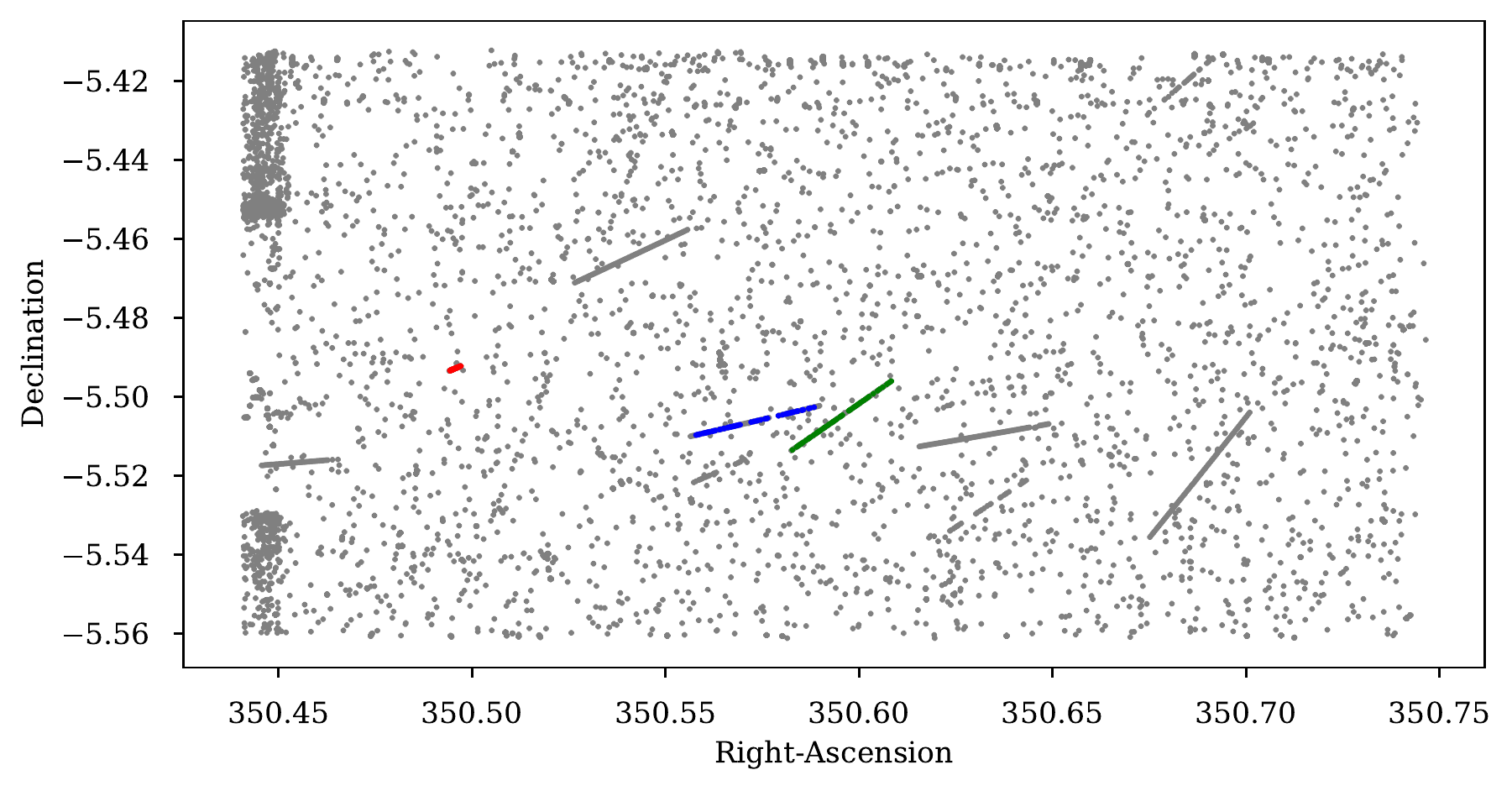}
\caption{DEEP transient catalog from one CCD. The red streak indicates one confrimed TNO within the image. The green streak shows a clean detection of an MBA. The blue pair of streaks were identified as separate by the Hough transform, but were linked by their slope and intercept during later processing steps.}
\label{fig:detected}
\end{figure}

As described above, there is a fair bit of artifacting in the transient catalogs, and many of the less scattered artifacts will be identified as streaks by the Hough transform. This is by design: the algorithm has very tunable thresholds for many parameters including line length, line density, linearity, and others and constraining these parameters too aggressively will limit spurious detections, but will also reduce the success rate in detecting true moving TNOs. Instead, we are fairly conservative with our thresholds to maximize our sensitivity to moving sources, true or spurious. The expected result of this is the detection of several true TNOs, but also several false detections. To eliminate the first batch of false detections, all streaks with slopes $>$ 10 are rejected as nonphysical, as are line segments with lengths of zero. In the case that some streaks may be broken, nearby line segments with the same slope and intercept are combined (Figure \ref{fig:detected}). Because of the probabilistic nature of the Hough method used, duplicate detections are also common, and are rejected as well.

\begin{figure}
\includegraphics[width=\textwidth]{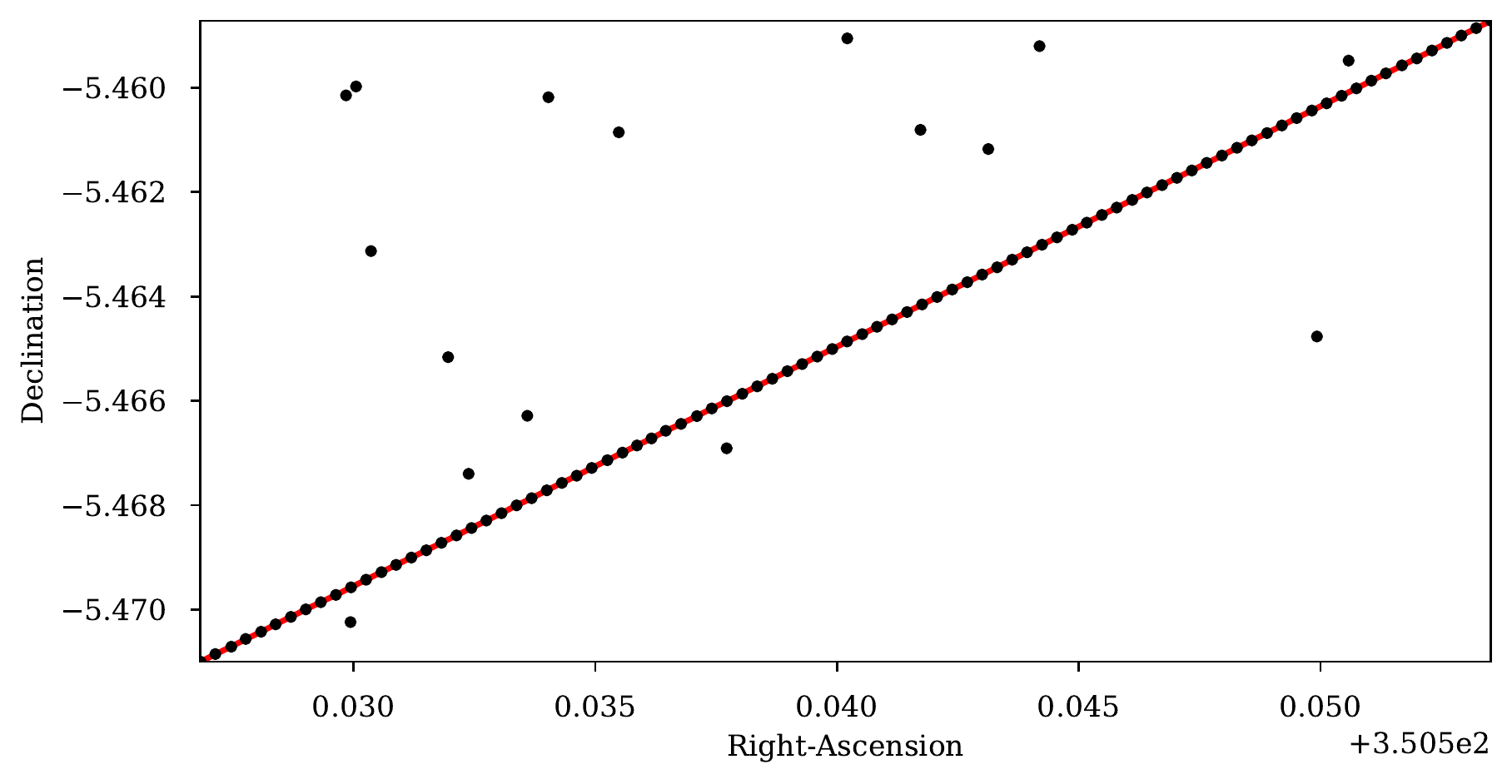}
\caption{Red: the line segment identified by the Hough transform algorithm. Black: the sub-set of transient sources from the DEEP catalog which is compared against the line segment.}
\label{fig:subframe}
\end{figure}

At this point, we expect that the majority of the points within the transient catalogs which fall within some threshold distance from a given line will correspond to that line. The set of points which correspond to a single line make up a full detection of a moving object. To avoid comparing every source in the catalog with every line, we sub-sample the catalog such that each line is only compared against points which fall within a rectangle bisected by that line (Figure \ref{fig:subframe}. This significantly reduces the computational demand of this step. Each source which appears more than 10$^{04}$ degrees from the line is rejected, and the remaining sources are sorted by the time of their respective exposures. 

Because each source should only appear once per exposure, each streak is searched for time-duplicate sources. For each set of duplicates, we compare their magnitudes against those of their nearest neighbors. Only the best-matching source is retained, while all other duplicates are rejected. The streaks are then checked for linearity in three dimensions: Right-Ascension, Declination, and time. The metric we use to evaluate linearity is the Pearson correlation coefficient (R), a value in the range (-1,1) which indicates the strength of a linear correlation between two variables \citep{Pearson1895}. A linear least squares regression is performed in the coordinate pairs of (RA,Dec), (RA,time), (Dec, time), and all streaks with R$<$ 0.9 are rejected. The remaining sets of points are the best candidates for real detected objects. 

Up to this point, we have not taken any steps to differentiate between different solar system objects. The relaxed constraints used in the streak detection were not designed to reject all MBAs or NEOs. This is by design - there is a considerable amount of science to be done with these other populations of objects, but that is beyond the scope of this paper, and will be reported in forthcoming work. 

Due to four-hour time-baseline for our astrometry, the orbital constraints on DEEP TNOs are not very strong. For differentiating between major populations, rate-of-motion is a reasonable proxy for heliocentric distance. We reject all rates outside of the range 1-6 arcseconds per hour, and consider the remaining set to be a relatively pure sample of TNOs. As expected, when filtered by rate-of-motion, the vast majority of the streaked sources in the DEEP images are MBAs, and only a small fraction have rates of motion consistent with TNOs. Because Centaurs are relatively rare, we do not see any such objects within our small sample.

\begin{figure}[htbp]
    \centering
    \begin{minipage}{0.2\linewidth}
        \centering
        \includegraphics[width=\linewidth]{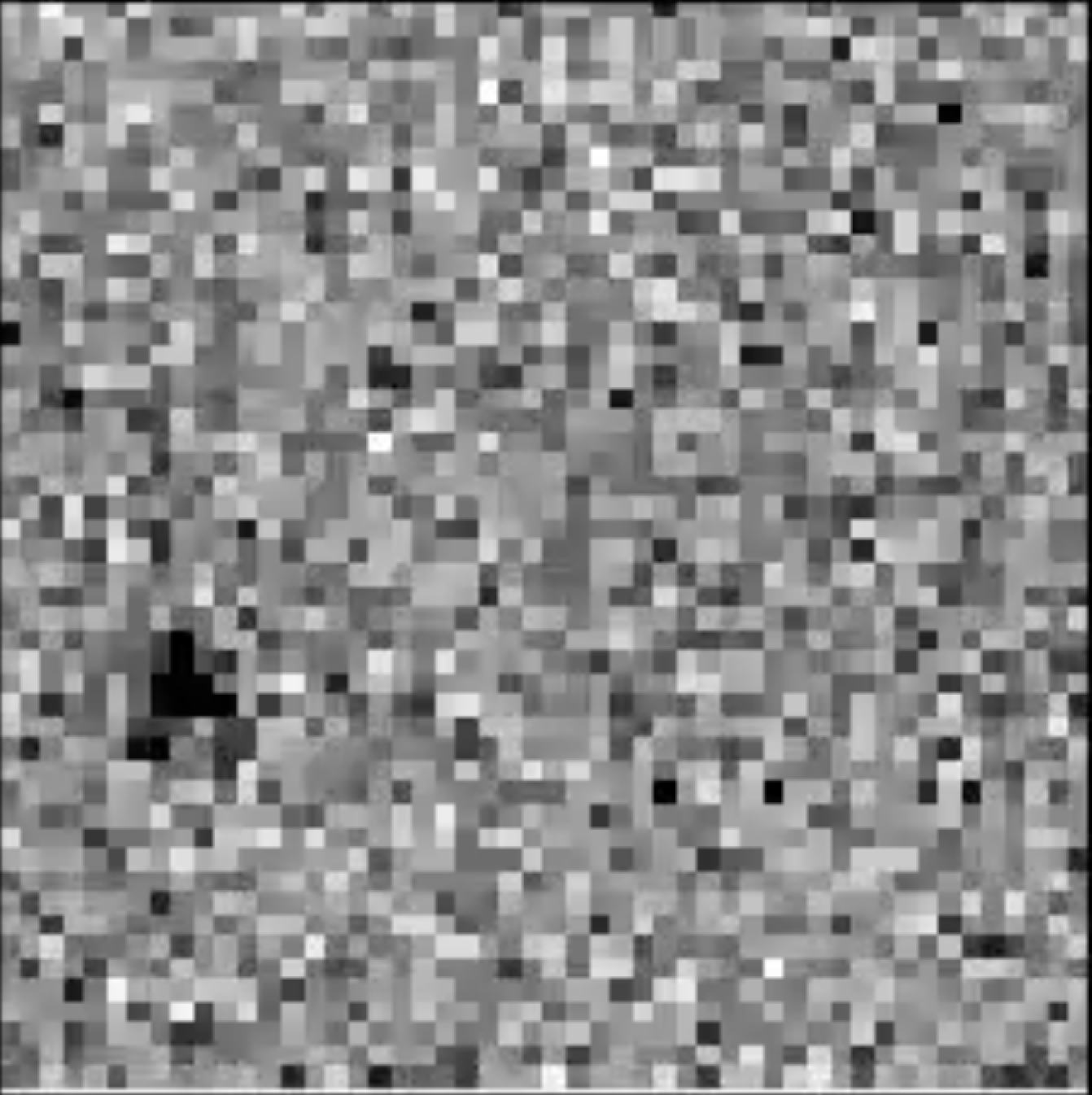}
        \label{fig:panel-a}
    \end{minipage}
    \hspace{0.02\linewidth}
    \begin{minipage}{0.2\linewidth}
        \centering
        \includegraphics[width=\linewidth]{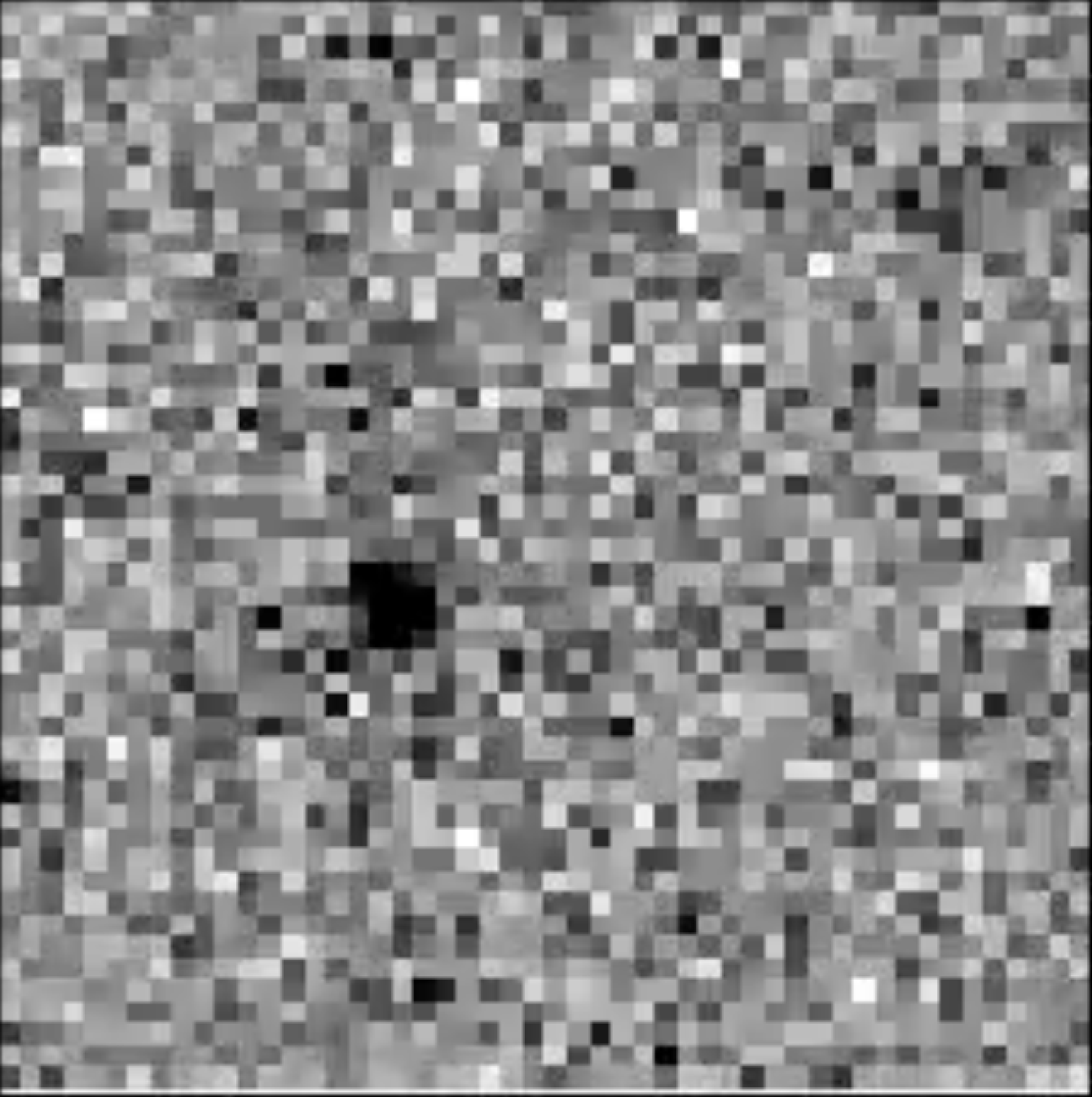}
        \label{fig:panel-b}
    \end{minipage}
    \hspace{0.02\linewidth}
    \begin{minipage}{0.2\linewidth}
        \centering
        \includegraphics[width=\linewidth]{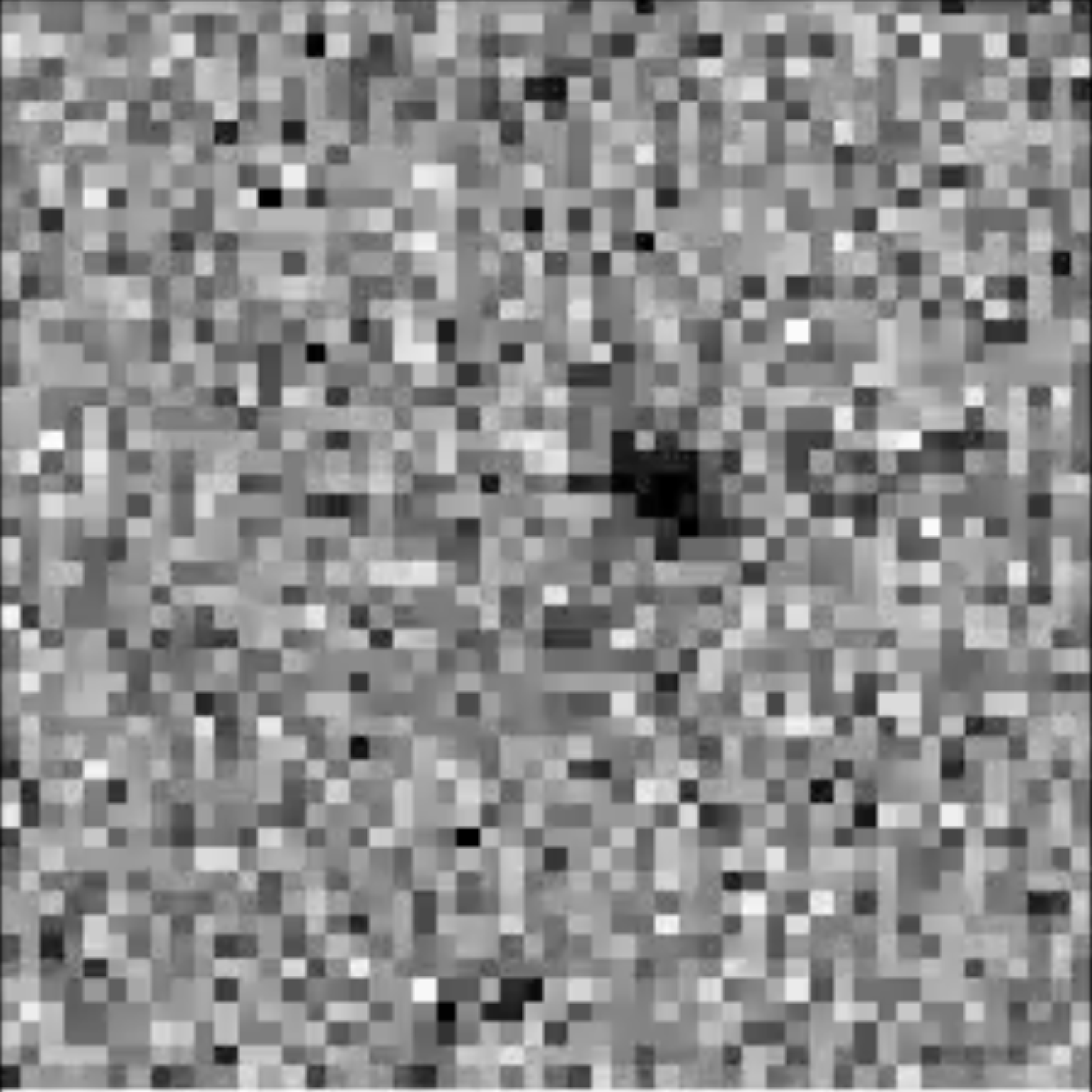}
        \label{fig:panel-c}
    \end{minipage}
    \hspace{0.02\linewidth}
    \begin{minipage}{0.2\linewidth}
        \centering
        \includegraphics[width=\linewidth]{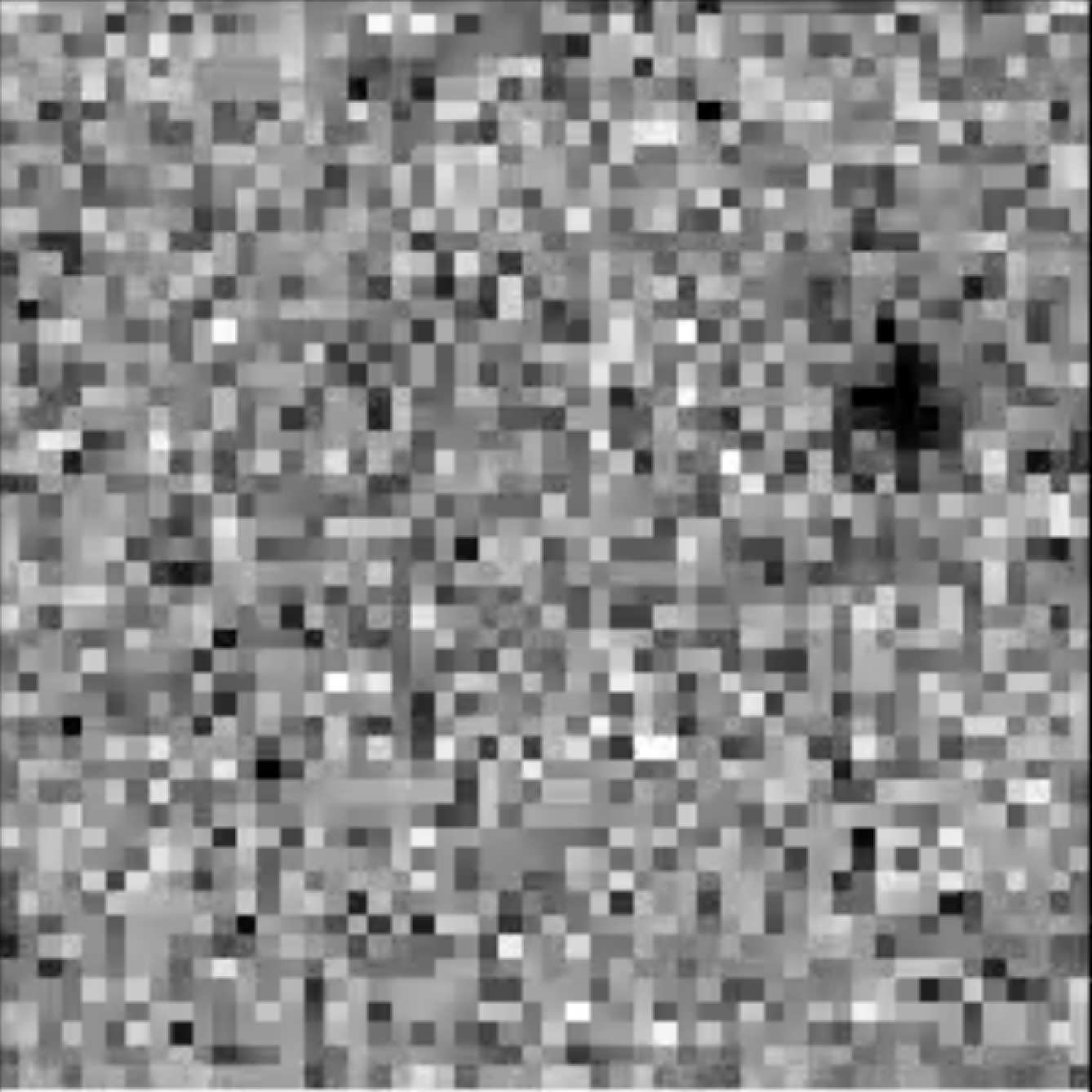}
        \label{fig:panel-d}
    \end{minipage}
\\
    \begin{minipage}{0.2\linewidth}
        \centering
        \includegraphics[width=\linewidth]{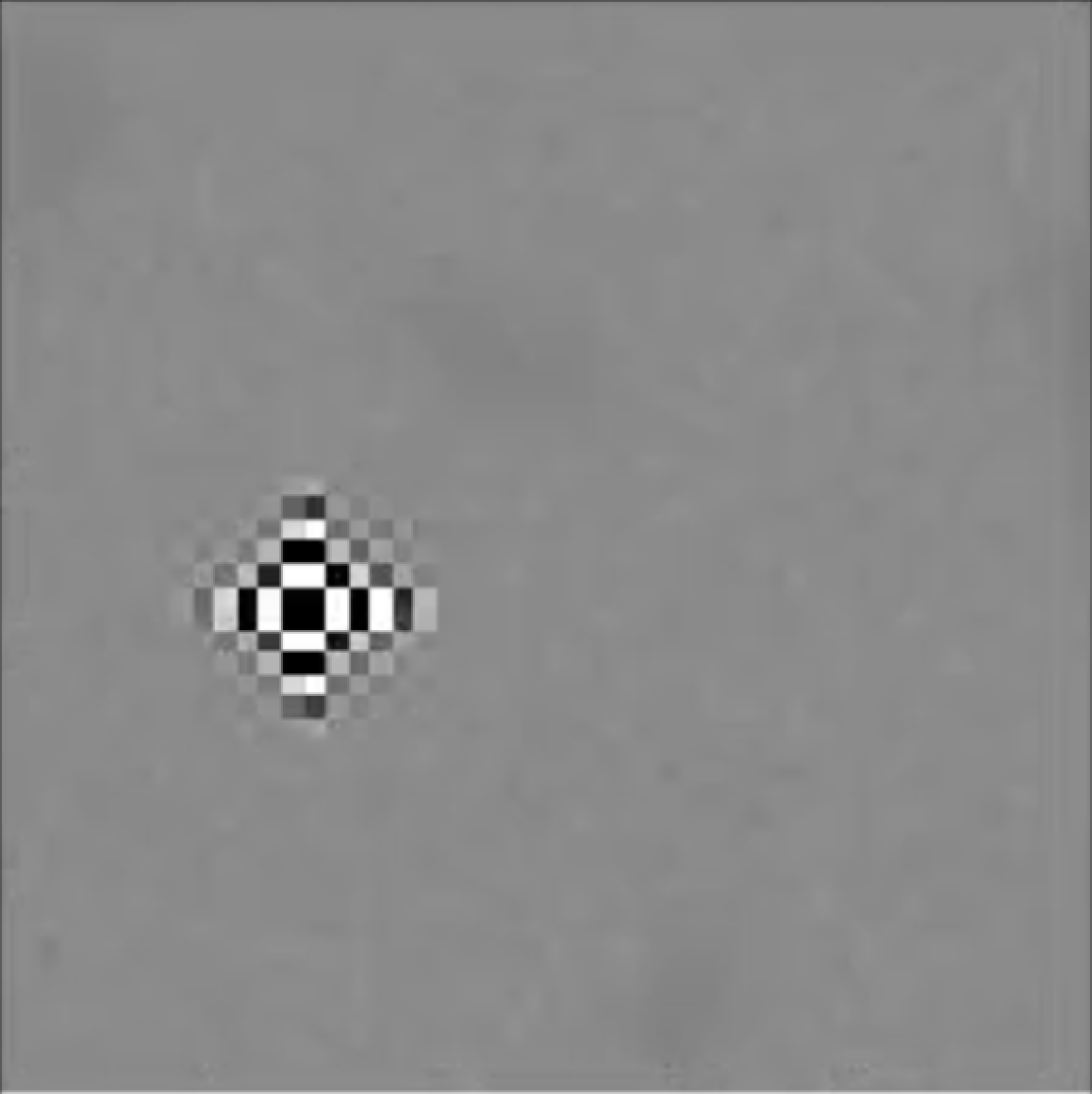}
        \label{fig:panel-e}
    \end{minipage}
    \hspace{0.02\linewidth}
    \begin{minipage}{0.2\linewidth}
        \centering
        \includegraphics[width=\linewidth]{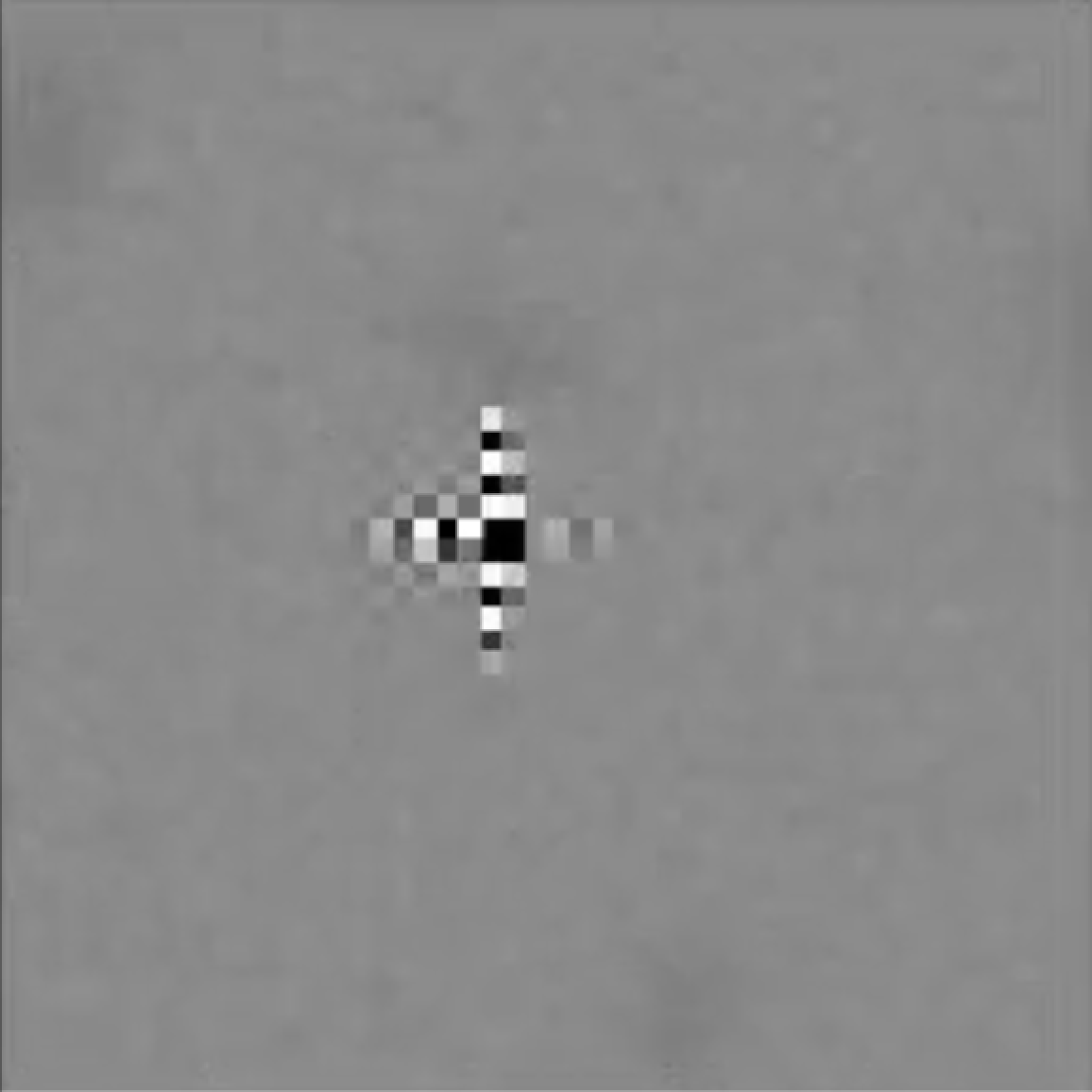}
        \label{fig:panel-f}
    \end{minipage}
    \hspace{0.02\linewidth}
    \begin{minipage}{0.2\linewidth}
        \centering
        \includegraphics[width=\linewidth]{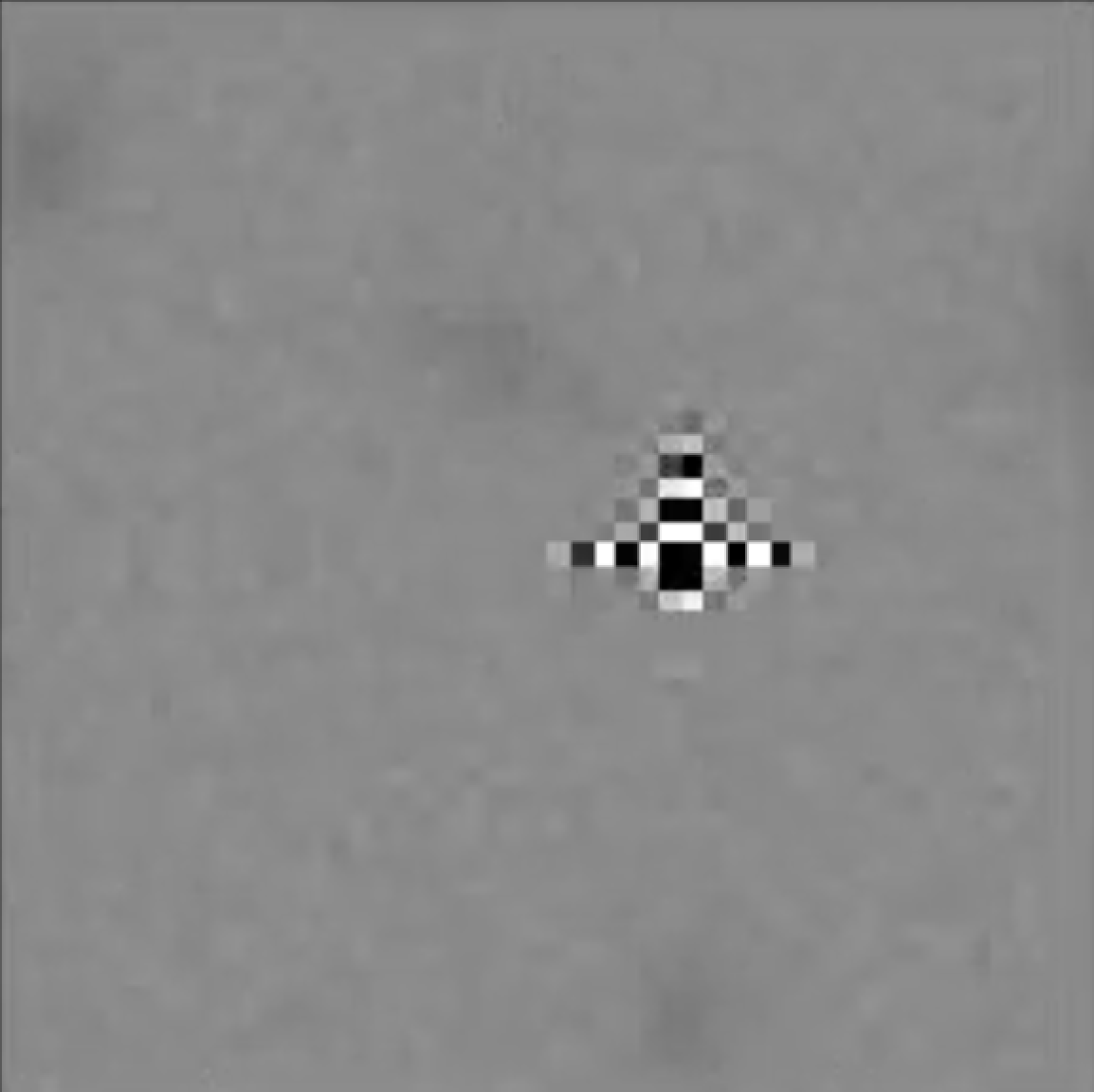}
        \label{fig:panel-g}
    \end{minipage}
    \hspace{0.02\linewidth}
    \begin{minipage}{0.2\linewidth}
        \centering
        \includegraphics[width=\linewidth]{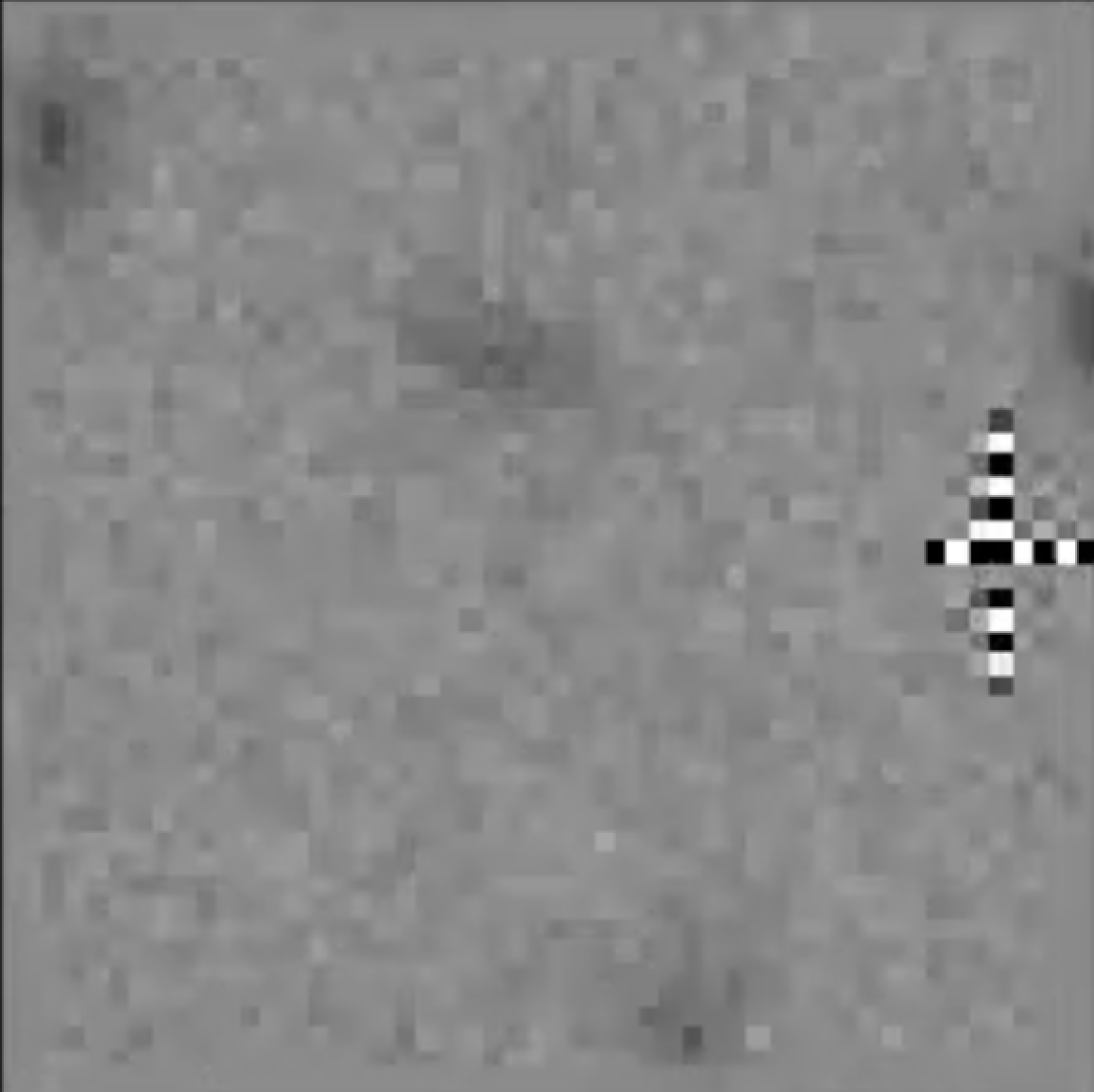}
        \label{fig:panel-h}
    \end{minipage}
    \caption{Top: Four frames from an animated thumbnail of images taken from a DEEP long stare. The center of the frame is stationary with respect to the celestial coordinates. The display lookup table is inverted such that the source appears darker. The confirmed TNO appears in strong relief and is clearly moving within the frame. Bottom: Four frames from a similar thumbnail from the same long stare. The moving point is clearly an artifact from a bad pixel drifting across the sky, rather than a true source}.
    \label{fig:realtno}
\end{figure}

Because the number of TNO candidates is small (on the order of 100), the next step of vetting the candidates can be performed manually. This is the final data processing step before we have a catalog of confirmed TNO detections.  While the majority of our data processing is performed on a catalog, the manual vetting involves visually inspecting the original images obtained with DECam. We generate animated thumbnails of the 50x50 pixel square surrounding a detected streak. When viewed in motion, the true detections are extremely apparent as moving sources tracking across the sky (Figure \ref{fig:realtno}), while false detections are easy to identify and reject. Once this visual vetting step is complete, our final product is a catalog of confirmed TNO detections, with four-hour photometric and astrometric measurements for each object. 

%%%%%%%%%%%%%%%%%%%%%%%%%%%%%%%%%%%%%%%%%%%%%%%%%%%%%%%%%%%%%%%%%%%%%%%%%%%%%%%%%%%%%

\section{Results} \label{sec:results}

\begin{figure}
\includegraphics[width = 0.7\linewidth]{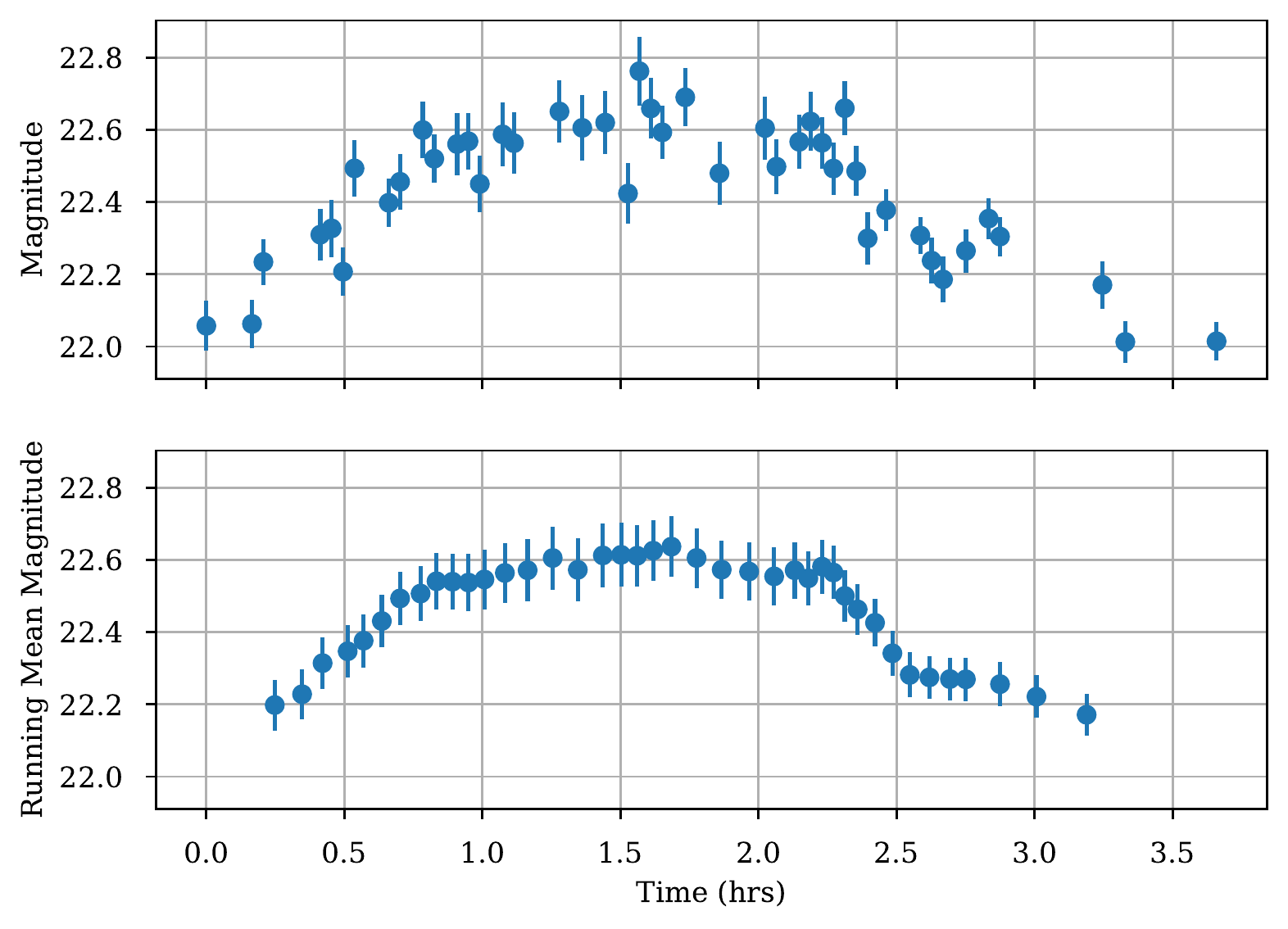}
\caption{Top: example of a typical DEEP TNO lightcurve for an object detected by our pipeline. Bottom: The same DEEP TNO lightcurve, with a running box mean applied in which each point is assigned the average value of itself and its four nearest neighbors. The random scatter in the top image seems to effectively disappear into the signal, reducing the overall overestimate given by computing a simple maximum-minimum amplitude. At the timescale of a typical TNO lightcurve such as this one, we can see that the more extreme spikes in the brightness are effectively smoothed out, while the overall lightcurve shape is preserved.}
\label{fig:boxmean}
\end{figure}

While a number of individual lightcurves are compelling on their own, many do not have clear structure from a by-eye examination. Instead, to investigate the shape distribution of our detected objects, we investigate the distribution of lightcurve amplitudes within our data. We start by computing a raw amplitude based on the maximum difference between the brightest and faintest photometric data points. This raw min-max amplitude is, however, artificially inflated due to the random scatter from the photometric error. To compensate for this inflation and crush away the random scatter, a running box mean with five-point bins was performed on the lightcurves. This does not perfectly remove the spread from the scatter, but is acceptable for our purposes (Figure \ref{fig:boxmean}).

\begin{figure}
\includegraphics[width=0.5\linewidth]{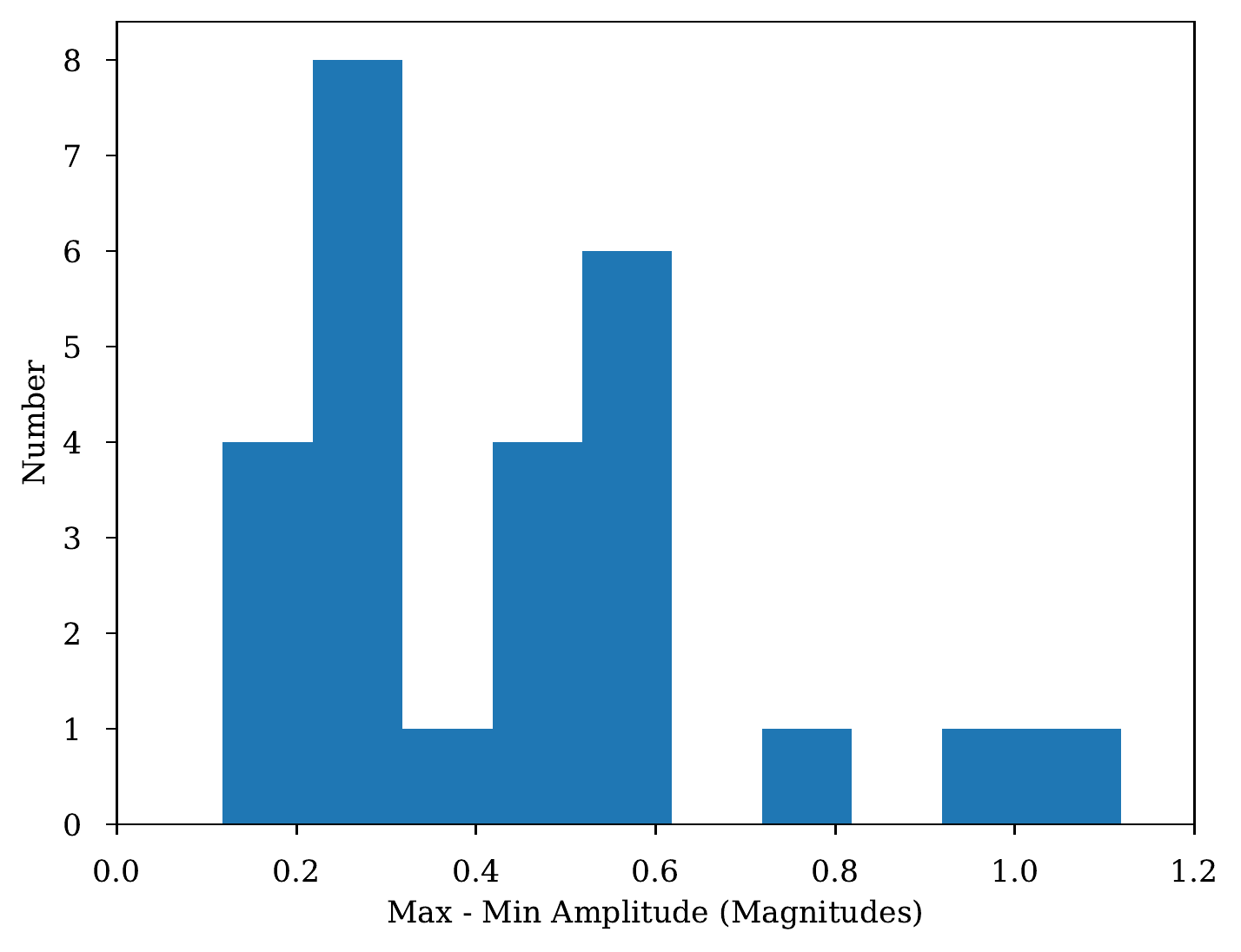}
\includegraphics[width=0.5\linewidth]{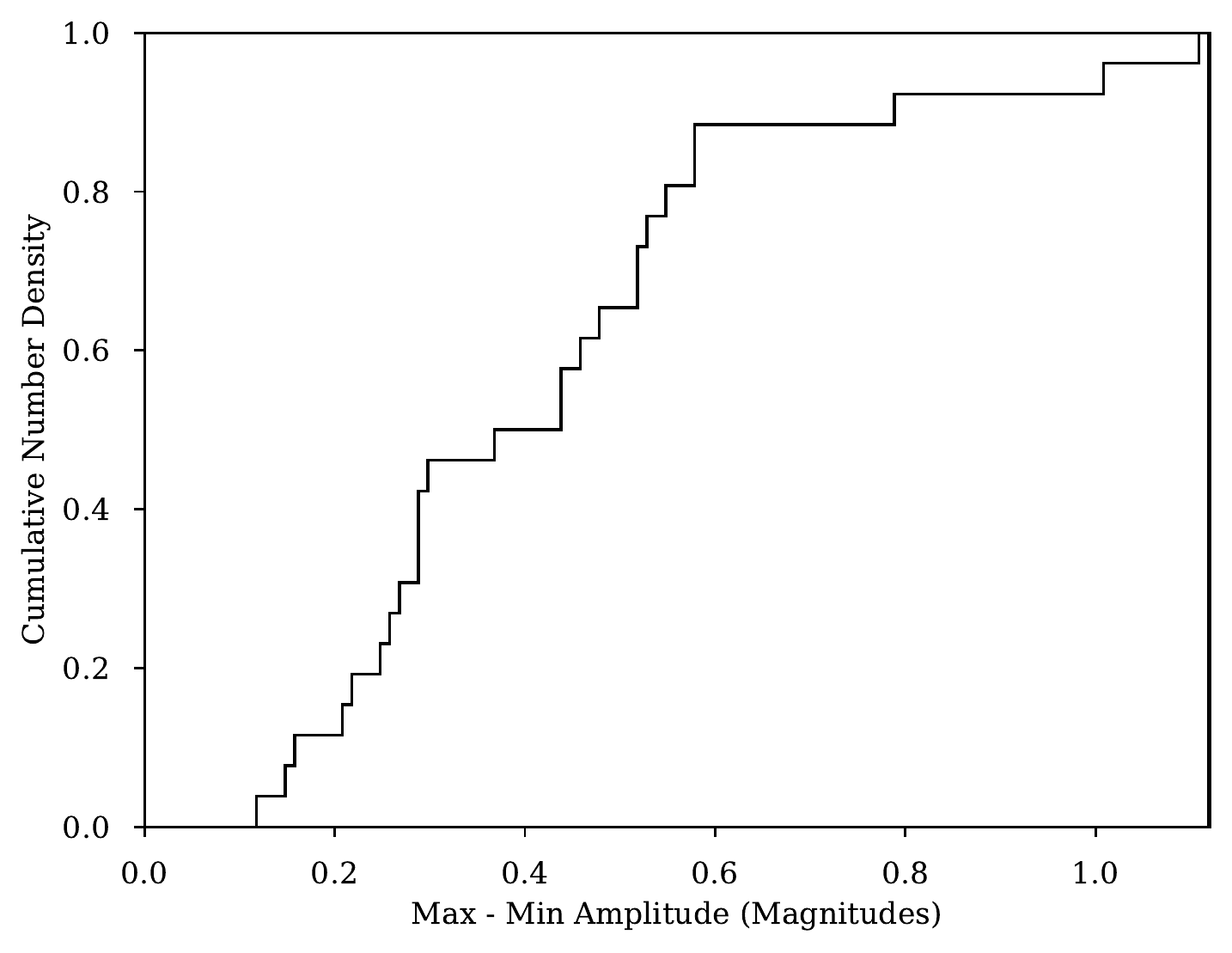}
\caption{\textit{Left:} Differential max-min partial lightcurve amplitude distribution for large (D$\sim$ 30-100km) DEEP TNOs. \textit{Right:} Cumulative density distribution for partial lightcurve amplitudes for DEEP single-exposure TNOs. The lack of objects in the lowest-amplitude bin is an effect of the photometric uncertainty in the data - particularly at the detection limit, even a spherical object with a homogeneous surface will exhibit some variation in its lightcurve.}
\label{fig:amp_distro}
\end{figure}

\citet{Sheppard2008} notes that the average TNO rotation period is roughly 8 hours, which exceeds the 4-hour photometric baseline of our DEEP long stares. So, for the majorxity of our objects, we do not expect to sample a full lightcurve period. Depending on the phase at which the partial lightcurve is sampled, the computed max-min amplitude will be less than the actual lightcurve amplitude. For a sinusoidal lightcurve, at least half of the rotation period must be sampled to guarantee that both a maximum and a minimum are present in the data, otherwise the correction is highly dependent on the sampled lightcurve phase. For this reason, the amplitudes that we measure provide lower limits on the actual amplitude of the object lightcurves. The resulting max-min amplitude distribution for this work is shown on the left panel of Figure \ref{fig:amp_distro}. Due to the small sample size, it is much more useful to investigate the cumulative amplitude distribution for the DEEP Year 1 TNO, which we show on the right panel of Figure \ref{fig:amp_distro}. 

\section{Discussion}\label{sec:discuss}
%\subsection{Shape Distribution}

The lightcurve amplitude distribution shown in Figure \ref{fig:amp_distro} suggests some interesting things about the shape distribution of our discovered DEEP objects. Here, we see a peak near low lightcurve amplitudes, with an envelope that gradually decreases towards higher amplitudes up to 1.1 magnitudes. We emphasize that these lightcurve amplitude values are nominally underestimates; we are not necessarily sampling full lightcurve periods and may not be capturing the full min-max variation in the brightness. Additionally, the spin-poles of these objects are not known. An object with a spin pole pointing at the observer will have no measured lightcurve, regardless of its true shape or rotation period, and only a spin pole pointed perpendicular to the observer will permit measurement of the full lightcurve amplitude. 

To understand this amplitude distribution in the context of the shape distribution of the outer Solar System, we constructed a model Kuiper Belt with an adjustable shape distribution. The first model distribution we tested against was that of a Kuiper belt populated entirely by objects with the same shape as Arrokoth. For the purposes of this broad lightcurve amplitude study, it is not necessary to fully model the shapes the individual objects. We model each object in the population as a 3-dimensional rotating prolate ellipsoid, with axis lengths (a $>$ b = c). We can then describe the shape of a given object by a single value: the axis ratio b/a. A spherical object would have b/a = 1, while a more elliptical object might have b/a much less than one. We give the bulk population an average axis ratio b/a = 0.32, roughly the axis ratio of Arrokoth. Their spin poles were oriented randomly with respect to the observer, and their rotational periods were randomly scattered between 2 hours and 20 hours. These populations were modeled using the methods described by \citet{Leconte2011,Mommert2018,McNeill2019}. We then simulated observations of this model population with the same cadence and similar characteristic noise as the DEEP long-stare sequences. This accounts for the slight artificial lightcurve amplitude inflation we see in Figure \ref{fig:amp_distro} and explains the lack of zero-amplitude objects. The second model distribution was treated exactly the same way, except the population was given an average shape of b/a = 0.8, consistent with other MBA and NEO populations \citep{McNeill2019}.  

\begin{figure}

\includegraphics[width=0.5\linewidth]{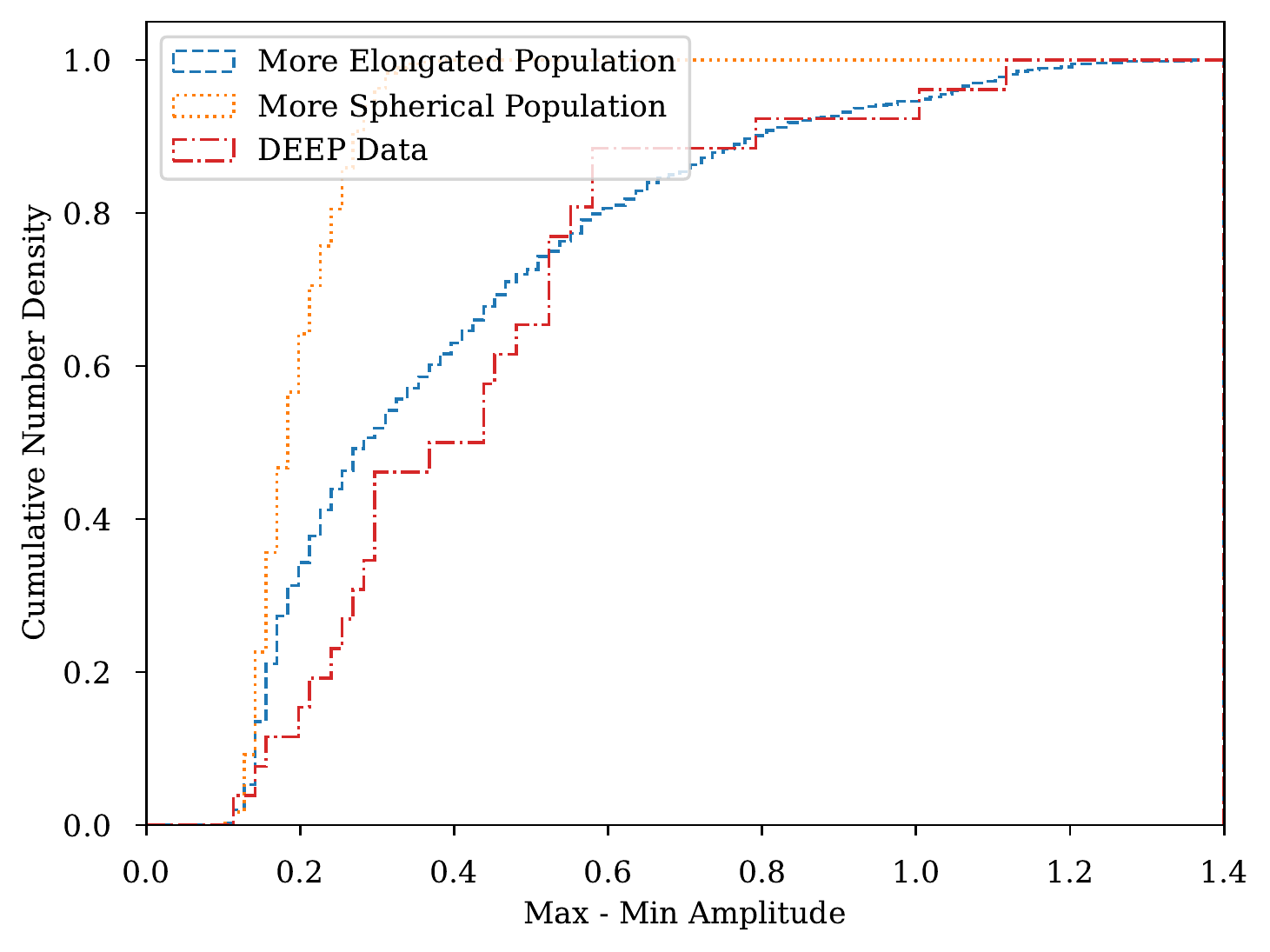}
\caption{Max-Min cumulative amplitude distributions for two sample model Kuiper Belt shape distruibutions as well as the cumulative distribution from the first year of DEEP (red). One model Kuiper belt shown here is given an average axis ratio of 0.32 (blue, consistent with the shape of Arrokoth). The other model Kuiper belt shown is given an average axis ratio of 0.8 (yellow), corresponding to a mostly spherical population. The DEEP data is apparently more consistent with a population of extremely high average ellipticity.}
\label{fig:amp_distro_model}
\end{figure}

The resulting model cumulative lightcurve amplitude distribution for a population consisting entirely of Arrokoth-shaped bodies, with an average axis ratio of b/a = 0.32 is shown on the left side of Figure \ref{fig:amp_distro_model}. The right panel of the same figure shows the model amplitude distribution for a population of much higher sphericity, with an average b/a of 0.8. Both panels have the cumulative amplitude distribution of the DEEP data plotted for comparison. Visual inspection clearly favors a very non-spherical population as the best match to the DEEP data. 

\begin{figure}

\includegraphics[width=0.5\linewidth]{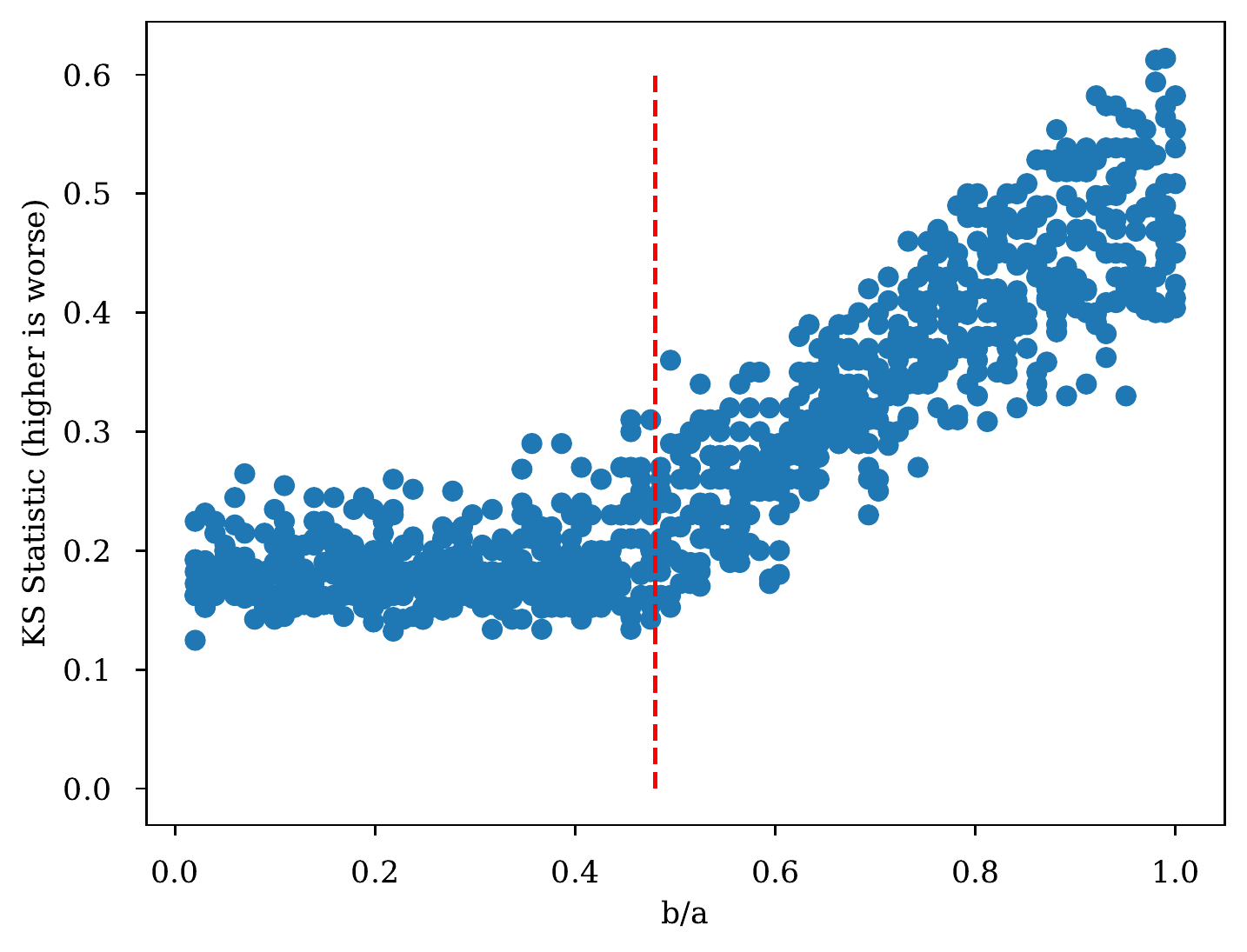}
\caption{Kolmogorov-Smirnov statistic values for simulated populations of varying average b/a axis ratios. A higher value for the K-S statistic hereimplies a higher confidence that the data does not match the model. The DEEP data is a good match to the model for average b/a values up to about 0.5 (dashed red line). As the axis ratio increases toward unity, the value of the K-S statistic increases, implying a poorer match to the data.}
\label{fig:boa_ks}
\end{figure}

By visual inspection, it is clear that the data is a poor match to the more spherical model population. To quantify this, we apply a monte carlo method to the forward model to vary the average shape and it's standard deviation for 1000 model populations. When we plot the Kolmogorov-Smirnov (K-S) statistic for each population as a function of b/a, the data take the form of a split-slope line \citep{Hodges1958}. A pair of linear regressions reveal the "knee" to be at b/a = 0.48, beyond which the K-S statistic rapidly increases (Figure \ref{fig:boa_ks}). By this metric, our data are most consistent with a population with an average ellipsoidal axis ratio of 0.5 or smaller: a very non-spherical population. 

We note that a single elongated object is not the only possible cause of rotational photometric variation. A contact binary will display similar variation with the same double-peaked lightcurve, but with a slightly different overall lightcurve shape \citep{SHOWALTER2021}. Another common source of photometric variation is surface heterogeneity - if the surface albedo changes, so will the brightness. The key difference here is that shape-driven variation will have a double-peaked lightcurve, while albedo-driven variation may only have one peak. 

\citet{Jeans1915,Chandrasekhar1963,LEONE1984} demonstrate that lightcurve amplitude in excess of 0.9 magnitudes is not possible for a strengthless rotating ellipsoid; unless there are additional sources of cohesive strength, at this point an object will succumb to rotational fissile instability and break apart. A close or contact binary, on the other hand, can easily store that angular momentum while remaining in a stable equilibrium. From this, it follows that a high fraction of high-amplitude lightcurves is also consistent with a high fraction of contact binaries. 

The presence of a high fraction of contact binaries in the Kuiper Belt is not without precedent. \citet{NESVORNY2019} suggests that hypothetical formation mechanisms such as the streaming instability may preferentially result in the formation of binaries rather than individual bodies. This is supported by observational evidence: \citet{NOLL2008} finds that approximately 30\% of low-inclination, 100km-class TNOs are separated similar-mass binaries, and \citet{Brown2006,WEAVER2006} note that binaries with smaller separations or larger mass disparities may be even more common. The most compelling evidence for the presence of contact binaries in the Kuiper Belt is the morphology of Arrokoth - the only sub-100km Kuiper Belt object ever visited by a spacecraft, Arrokoth is an excellent example of a contact binary \citep{STERN2019}. \citet{MCKINNON2020} describes the mechanism by which angular momentum might have been removed from the separated binary which might have evolved into the contact binary system we see at Arrokoth today. The occurrence of the same mechanism elsewhere in the trans-Neptunian region implies that the high observed fraction of separated binaries may correspond to a high fraction of contact binaries. 

A similar partial lightcurve amplitude analysis is presented in \citet{SHOWALTER2021}. They compile TNO lightcurve amplitudes from the literature and examine their cumulative amplitude distributions. One of the more puzzling results from their work is that, when separated by survey, the distribution of the data from the Outer Solar System Origins Survey (OSSOS) does not take the same shape as the rest of the sample \citep{Alexandersen_2019}. At a confidence of 99.9997\%, they find that the OSSOS sample is not drawn from the same underlying distribution as the rest. When comparing the DEEP data to these distributions, we find that our objects fall somewhere in between the two different distributions, i.e. we are more enhanced in high-amplidude objects than most of the literature, but slightly less so than OSSOS. They also note that the OSSOS objects follow a different trend in amplitude vs. H-magnitude than the other objects from the literature. 

We suggest a plausible explanation for this: The average absolute magnitude of the OSSOS catalog is H = 9, while the remaining population has an average H-magnitude of 6.5. The OSSOS objects, then, are nearly 10x smaller than the rest of the literature.  Our DEEP objects have an estimated average H-magnitude of 7.5, intermediate in size between OSSOS and the remainder of the literature. Both DEEP and OSSOS have, then, catalogued objects in a different size regime than the bulk of the literature, and have found their populations to be much less spherical. This is observational evidence that an average TNO's axis ratio may have a strong size dependence, with smaller objects being more elliptical than larger objects. This is consistent with arguments presented in \citep{Alexandersen_2019,BENECCHI2013,Audrey_2019, NESVORNY2019}. The \citet{SHOWALTER2021} result only implies a lack of linear correlation between size and lightcurve amplitude, not an overall lack of size-dependence. If this hypothesis holds, the ligthcurve amplitude distribution for the faintest shift-and-stack DEEP TNOs (forthcoming) will have an even more extreme spread than those measured by OSSOS.

While our ability to constrain the dynamical population of these objects is limited by our short observation arcs, we can make a couple of constraining assumptions based on the observing geometry. Most of the detected objects are found in our B1 fields, which are very close to Neptune's ecliptic longitude. This is a region where resonant objects spend very little time, as the most stable resonant objects are typically at periheleon during conjunction with Neptune. Our DEEP fields are also concentrated close to the ecliptic, which may contribute to low-inclination objects being over-represented. Because of these two factors, we expect the DEEP year 1 samplie is dominated by cold-classical TNOs . Thus, this work is also consistent with these in-situ formation models. 

\begin{figure}

\includegraphics[width=0.5\linewidth]{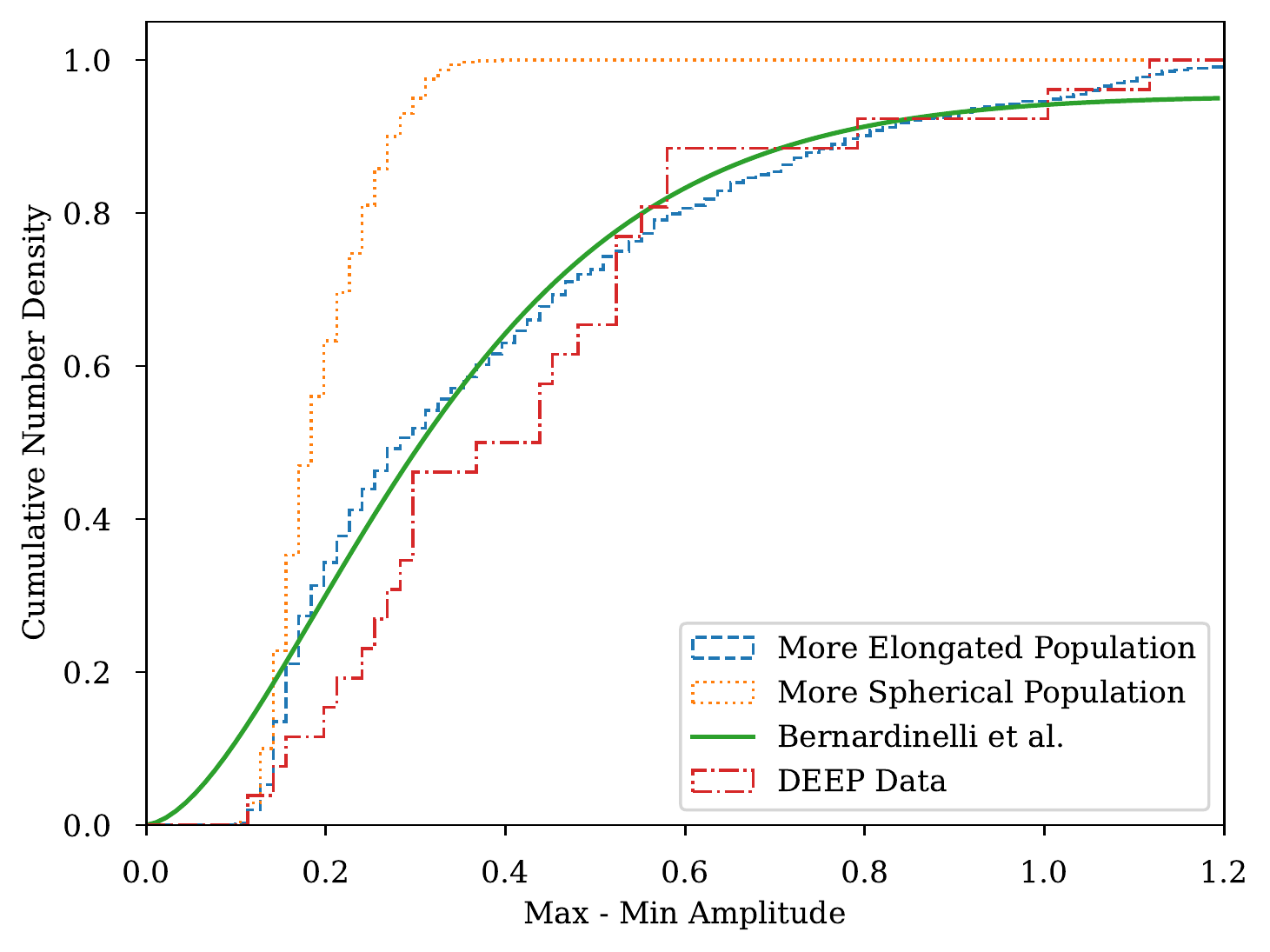}
\caption{Max-Min cumulative amplitude distribution for our DEEP sample, as well as the model distributions for both end-member populations (average b/a = 0.8 and 0.31, respectively, overplotted with the cumulative amplitude beta distribution of the from \citet{Bernardinelli2023}. The DEEP data and that from Bernardinelli et al. are each a good visual match to the elongated model population, and fit poorly to the more spherical population.} 
\label{fig:pedro_compare}
\end{figure}

In their Arxiv preprint, \citet{Bernardinelli2023} present photometry for a large number of TNOs, and examine the lightcurve variation as a function of dynamical class. They find the dynamically cold classical TNOs (non-resonant objects with low eccentricities and low inclinations) to have much stronger photometric variation than other populations, which is consistent with in-situ cold-classical formation models \citep{Nesvorny2010,Nesvorny2018,NESVORNY2019}. \citet{Bernardinelli2023} also present TNOs in the same size-regime as this work, with a similar sensitivity of VR~24, making theirs a very good sample to compare this work against. A comparison between our amplitude distribution and theirs is shown in Figure \ref{fig:pedro_compare}. We note that for both samples, $\sim$50\% of objects have lightcurve amplitudes of $\leq$ 0.4 magnitudes, and $\sim$80\% of objects have lightcurve amplitudes of $\leq$ 0.6 magnitudes, indicating at least a first-order agreement.

\citet{Audrey_2019} discusses the number of contact binaries within the Cold Classical Kuiper Belt. Their analysis, like ours, examines the distribution of partial lightcurve amplitudes for a number of TNOs. Our objects are, on average, about a magnitude fainter, but because our sample is probably dominated by Cold Classical TNOs, their work is a reasonable comparison to ours. Our distribution is broader than that presented in the aforementioned work; with a slightly less spherical average shape. One hypothesis we posit for this difference is that while the DEEP survey is an un-targeted discovery survey, many of the surveys addressed by \citet{Audrey_2019} are targeted surveys of known objects. This introduces a secondary discovery bias which may trend towards rounder objects, as high-amplitude objects may spend more of their time below a given faint detection threshold than a similarly sized spherical object. 

\section{Conclusions and Future Work} \label{sec:conclusions}

In this work, we discuss the identification and analysis of single-exposure TNOs from the DEEP survey. The novel Hough transform computer vision technique we use for identifying Solar System transients is presented in depth, along with a subsequent visual vetting to confirm detected TNOs. Each detected TNO has a 4-hour lightcurve at a 2-minute cadence, which allows us to evaluate the partial rotational lightcurve morphologies. The distribution of (max-min) lightcurve amplitudes provides constraints on the shape distribution of our detected TNOs. Based on a comparison to several model amplitude distributions, obtained by simulating observations of model populations, we find our DEEP data to be most consistent with an average axis ratio of b/a $\leq$ 0.5. 

We also discuss the interpretation of our data with respect the contact binary incidence rate of the Kuiper belt. We suggest that our data is consistent with a high fraction of contact binarity for 100-km TNOs; this hypothesis is well-supported by other theoretical and observational work in the literature. Our result provides an observational constraint on existing formation models such as those presented by \citet{NESVORNY2019,MCKINNON2020}.

This work only addresses the first year (2019) of survey data. Future work will involve performing a similar study on objects from subsequent observing epochs, including multi-year linking (discussed in part in \citet{Smotherman_2023}). This will also allow us to impose stronger constraints on the orbits of our discovered TNOs. Because objects at 40AU only move a few arcseconds per hour on the sky, we expect to have re-observe the same objects during subsequent years, which will permit orbit fitting and a more powerful dynamical constraint. This multi-epoch work is beyond the scope of this paper. We also only present here those objects appearing at $>$5 sigma in the single DEEP exposure; binned photometry from fainter objects detected with digital tracking at $<$5 sigma \citep{Napier_2023} will allow us to further expand our sample size and improve the lightcurve amplitude distribution presented here.

\acknowledgments

This work is based in part on observations at Cerro Tololo Inter-American Observatory at NSF’s NOIRLab (NOIRLab Prop. ID 2019A-0337; PI: D. Trilling), which is managed by the Association of Universities for Research in Astronomy (AURA) under a cooperative agreement with the National Science Foundation. We thank the CTIO staff for their fantastic support during our observations. 

This work is supported by the National Aeronautics and Space Administration under grant No.\ NNX17AF21G issued through the SSO Planetary Astronomy Program and by the National Science Foundation under grants No.\ AST-2009096 and AST-1409547. This research was supported in part through computational resources and services provided by Advanced Research Computing at the University of Michigan, Ann Arbor. This work used the Extreme Science and Engineering Discovery Environment \citep[XSEDE; ][]{XSEDE}, which is supported by National Science Foundation grant number ACI-1548562. This work used the XSEDE Bridges GPU and Bridges-2 GPU-AI at the  Pittsburgh Supercomputing Center through allocation TG-AST200009.

H. Smotherman acknowledges support by NASA under grant No.\ 80NSSC21K1528 (FINESST). H. Smotherman, Mario Juri\'c and P. Bernardinelli acknowledge the support from the University of Washington College of Arts and Sciences, Department of Astronomy, and the DiRAC Institute. The DiRAC Institute is supported through generous gifts from the Charles and Lisa Simonyi Fund for Arts and Sciences and the Washington Research Foundation. M. Juri\'{c} wishes to acknowledge the support of the Washington Research Foundation Data Science Term Chair fund, and the University of Washington Provost’s Initiative in Data-Intensive Discovery.

We would like to thank an anonymous referee for providing valuable comments which have considerably improved the overall quality of the paper.






\bibliography{DEEPIV}{}
\bibliographystyle{aasjournal}



\end{document}